%% file: paper.tex
\documentclass[sigconf,anonymous=false]{acmart}

\usepackage{booktabs} % For formal tables
\usepackage{flushend}

\definecolor{hotpink}{RGB}{235,0,65}

\setcopyright{rightsretained}
\usepackage{subfigure}

\acmDOI{10.475/123_4}

\acmISBN{123-4567-24-567/08/06}

\acmConference[SIGIR'18]{ACM SIGIR}{July 2018}{Michigan USA}
\acmYear{1997}
\copyrightyear{2018}

\acmArticle{4}
\acmPrice{15.00}

\begin{document}
\title{Human Perception of Surprise: A User Study}
%\subtitle{Extended Abstract}

\author{Nalin Chhibber}
\orcid{1234-5678-9012}
\affiliation{%
  \institution{University of Waterloo}
  \city{Waterloo}
  \state{Ontario, Canada}
}
\email{nalin.chhibber@uwaterloo.ca}

\author{Rohail Syed}
\affiliation{%
  \institution{University of Michigan}
  \city{Ann Arbor}
  \state{Michigan, USA}
}
\email{rmsyed@umich.edu}

\author{Mengqiu Teng}
\affiliation{%
  \institution{University of Michigan}
  \city{Ann Arbor}
  \state{Michigan, USA}
}
\email{mengqiu@umich.edu}

\author{Joslin Goh}
\affiliation{%
  \institution{University of Waterloo}
  \city{Waterloo}
  \state{Ontario, Canada}
}
\email{jtcgoh@uwaterloo.ca}

\author{Kevyn Collins-Thompson}
\affiliation{%
  \institution{University of Michigan}
  \city{Ann Arbor}
  \state{Michigan, USA}
}
\email{kevynct@umich.edu}

\author{Edith Law}
\affiliation{%
  \institution{University of Waterloo}
  \city{Waterloo}
  \state{Ontario, Canada}
}
\email{edith.law@uwaterloo.ca}

%\subtitle{Extended Abstract}

\begin{abstract}
	Understanding how to engage users is a critical question in many applications. Previous research has shown that unexpected or astonishing events can attract user attention, leading to positive outcomes such as engagement and learning. In this work, we investigate the similarity and differences in how people and algorithms rank the surprisingness of facts.  Our crowdsourcing study, involving 106 participants, shows that computational models of surprise can be used to artificially induce surprise in humans. 
\end{abstract}


\maketitle

\input{content}

\bibliographystyle{ACM-Reference-Format}
\bibliography{references}

\end{document}

%% file: content.tex
\section{Introduction}
	Nobel laureate Daniel Kahneman \cite{kahneman2011thinking} described the human mind as a combination of fast and slow thinking systems. 
	%System 1 operates automatically and quickly with little or no effort and System 2 allocates attention to the mental activities that demand effort. 
	%His work focused on interactions between these systems and the problems inherent in those interactions.
	The slow thinking system, which allocates attention to the mental activities that demand effort, is inherently lazy; and if the tasks at hand do not demand immediate attention, our brain delegates the mental legwork to the fast thinking system, which operates automatically and quickly with little to no effort. This implies that people pay attention to activities that fit a pattern they recognize, but they can also rapidly lose focus if their attention is not sustained. Thus, it may be important to periodically provide information that is not only useful, but also unexpected, in order to keep users attentive and engaged.  
	
	Our motivation is to develop methods that can eventually be used in educational technologies to enhance student engagement.  Prior work \cite{LawRecyclo2017} has shown that surprise or violation of expectations can be used to engage users by motivating them to seek information about certain known unknowns.   At the same time, researchers have shown that surprise can have positive effects on memory formation \cite{ranganath2003cognitive, hasselmo1999neuromodulation, wallenstein1998hippocampus,Parzuchowski2008}, and associative learning \cite{schultz2000neuronal, fletcher2001responses}.

l	Surprisingness of a topic has been studied under both subjective (user-driven) \cite{liu1997using, reisenzein2000subjective} and objective (data-driven) measures \cite{freitas1998objective}. In this work, we investigate algorithms for extracting and ranking surprising facts about a given topic.  We evaluate these techniques by running a study on Amazon Mechanical Turk, where we ask participants to rate the surprisingness of the extracted facts, compare their ranking to algorithmic rankings, and also ask them to explain what makes a fact surprising or not surprising.  The rest of this paper describes our primary research questions, system description and evaluation methodology, followed by a discussion of design implications.

\section{Related Work}
	
%\subsection{Quantifying Surprise}
	Although surprise has long been considered a subjective quantity primarily based on or influenced by personal feelings, tastes, or opinions \cite{reisenzein2000subjective}, there is significant work on quantifying surprise as an objective measure. Existing techniques can roughly be classified into the following two categories.
	%: \emph{Shannon surprise}
	%that treats surprisingness as the log-likelihood of a single data point given a statistical model of the world;
	%and \emph{Bayesian models of surprise}
	%which define surprisingness in terms of changes in model parameters induced by a new data point.
	\\\\
	%\begin{figure*}[h!]
	%	\centering
	%	\includegraphics[height=135px]{figures/pipeline.png}\hfill
	%	\caption{Three phase pipeline of extracting, validating and finally using surprise in human-ai interactions}
	%\end{figure*}
	% https://plus.maths.org/content/information-surprise
	\textbf{Shannon Surprise:} %Shannon Surprise is the log-likelihood of a single data point given a statistical model of the world. This idea originated from Shannon's interest in measuring the information contained in a message \cite{shannon1948mathematical}.
	%A naive technique would be to simply count the number of words contained in the message, but it would give the sentence \textit{"The sun is the center of our solar system"} the same information value as the sentence \textit{"The sun is classified as yellow dwarf star"} when the second is clearly more informative than the first. To overcome this, we can assign weight to words, e.g., to decrease the importance of words such as articles (like "a", "an" or "the") and pronouns (like "he", "she" or "it") that do not contribute much to the semantics of a message.  In contrast, 
	Shannon Surprise \cite{shannon1948mathematical} uses information content as a measure of surprise, based on the frequency with which words appear in a language model. Uncommon words occurring together are more likely to be informative and induce higher levels of surprise.
	%such as "dwarf" or "star", 
    Surprisingness $s(w)$ associated with a single word $w$ can be represented as the reciprocal of its probability of occurrence $p(w)$, so that $s(w) = 1/p(w)$. This formulation can be extended to a complete sentence by summing over the logarithms of the reciprocal probabilities of all words in the sentence. Thus, surprisingness of a statement containing $N$ words \{$w_1$, \ldots, $w_N$\} with corresponding probabilities \{$p_1$, \ldots, $p_N$\} can be computed as:
%	$$ S = w_1 \log {\left(1/w_1\right)} + w_2 \log {\left(1/w_2\right)} + ... + w_n \log {\left(1/w_ n\right)}$$
%	or simply
\begin{align*}
	S &= -~(p_1 \log {\left(p_1\right)} + p_2 \log {\left(p_2\right)} + \ldots + p_N \log {\left(p_N\right)})
\end{align*}
	This definition of surprise is similar to the calculation of entropy which gives a logarithmic measure of the number of states with significant probability of being occupied. Therefore, according to Shannon theory, unlikely events are more likely to be surprising.
	
	Inspired by Shannon's definition of surprise, Lin et al. \cite{lin2009identifying} used the ratio of assertion frequency to object frequency to find distinguishing assertions.
	%which is similar to existing notions of surprisingness and unexpectedness.
	They defined {\it assertion frequency} as the total number of times an assertion occurs and {\it object frequency} as the number of times the object appears in a sample of ten million random TextRunner assertions. Exman et al.\cite{exman2014interestingness} broadly defined interestingness as a composition of relevance (to a domain area) and unexpectedness. In a similar manner, Tsurel et. al. defined surprisingness of an article as the inverse of the average similarity to other articles in a category \cite{tsurel2017fun}.
	\\\\
    \textbf{Bayesian Surprise:}
    The problem with Shannon surprise is that it uses `unlikely' events as a proxy for `unexpected' events. In real life, an event can be unlikely without being unexpected: for example, the possibility of seeing a black cat sitting on a white bench is unlikely but not unexpected, and hence not surprising.
%	We argue that such metrics alone are not sufficient to quantify unexpectedness of information because they
	Shannon surprise also does not account for prior beliefs of users and how people's baseline knowledge affects their perception of surprise. The Bayesian model of surprise handles such limitations by measuring the differences between posterior and prior beliefs of people and describes how data affects natural or artificial observers. The Bayesian definition of surprise gives a consistent formulation of computational surprise under minimal axiomatic assumptions.
	Like Shannon surprise, Bayesian surprise also considers surprise as an information-theoretic concept which can be derived from first principles and formalized analytically across spatio-temporal scales, sensory modalities, and more generally, data types and data sources. The Bayesian framework to quantify surprise uses probabilistic concepts to cope with uncertainty and considers prior and posterior distributions to capture subjective expectations.
	Itti and Baldi defined Bayesian surprise as the Kullback-Leibler (KL) divergence \cite{kullback1951information} between the prior $\pi_0(\theta)$ and the posterior $\pi(\theta|X)$ beliefs about the model parameters $\theta$ either in the form
	$S_{Bayes}(X;\pi_0)=D_{KL}[\pi_0(\theta)||\pi(\theta|X)]$, or in the mirror form $D_{KL}[\pi(\theta|X)||\pi_0(\theta)]$ \cite{itti2006bayesian}. Their previous work has explored the effect of Bayesian surprise on attention \cite{baldi2010bits}, but little is known about how these computational models of surprise work with human users.

\section{Surprise Extraction Algorithms}
For our study, we implemented two algorithms for retrieving surprising facts that represent different approaches to computational surprise modeling.\\
\indent \textbf{Technique 1} extracts and ranks surprising facts using the approach proposed by Tsurel et al. \cite{tsurel2017fun}, which uses Wikipedia to calculate how surprising it is for an article to belong to a category. %, based on relationship between articles.
The score for a particular article-category pair was based on \emph{surprise} and \emph{cohesiveness} scores. Surprise is measured as the average semantic dissimilarity between the article and all other articles in that category. Cohesiveness is measured on the category as the average similarity of articles within that category to each other. The premise is that the relation between two entities (article and category) would be surprising to people if they were semantically dissimilar but the category was not too broadly defined. In this study, we used their approach to find 10 science- or history-related topics with at least 5 article-category pairs%In this study, we used their approach to find 10 article-category pairs for science- and history- related topics
\footnote{We used the full Simple Wikipedia as our source of term-document frequency data.}.\\
\indent \textbf{Technique 2} extracts the interesting assertions from DBPedia Wikipedia infobox database similar to \cite{lin2009identifying}, and ranks them based on their surprisingness. It fetches the N-triples in the form of <subject, predicate, object> from DBPedia and removes the parts that do not relate to ontological details of the entity. In order to compare the effectiveness of both approaches in matching human's perception of surprise, we retained only the triples that were similar to the ones extracted from Technique 1. Then, we calculated how frequently the words from secondary entity (object) appear in the web within the context of the primary entity (subject). For this, our system crawled the web using Google's Custom Search API when the primary entity (chosen topic) is used as a search query. To reduce noise in the search data, we restricted our search to only consider pages from the Wikipedia domain. 
In order to calculate the surprisingness scores of triples, we multiplied the term frequency of secondary entities appearing in search result pages with their inverse document frequency (i.e logarithm of the total number of search pages returned/number of search pages containing the secondary entity). This is similar to the work by \cite{exman2014interestingness} which defines interestingness as a composition of relevance and unexpectedness and reflects how important a triple is to the primary entity from each Wikipedia article, amongst all other articles in the collection of search results.
% In order to calculate the surprisingness scores of the triples, we followed the approach similar to \cite{exman2014interestingness} using term-frequency and inverse document frequency, considering each search result as a separate document and set of all pages from search results as a collection. This technique reflected how important a triple is to the primary entity from each Wikipedia article, amongst all other articles in the collection of search results. 

\section{User Study}

Our objective is to understand how people perceive surprise in sentences, which factors (e.g., prior knowledge) influence the perception of surprise, and how the different algorithmic approaches compare in terms of their ability to extract facts that are actually perceived by people as surprising.   In addition, we are interested in capturing people's own rationale for what makes a fact surprising versus not surprising.

\begin{figure*}[t!]
\centering
\subfigure[Step 1: Users select a topic of their interest.]{
\fbox{\includegraphics[height=4cm]{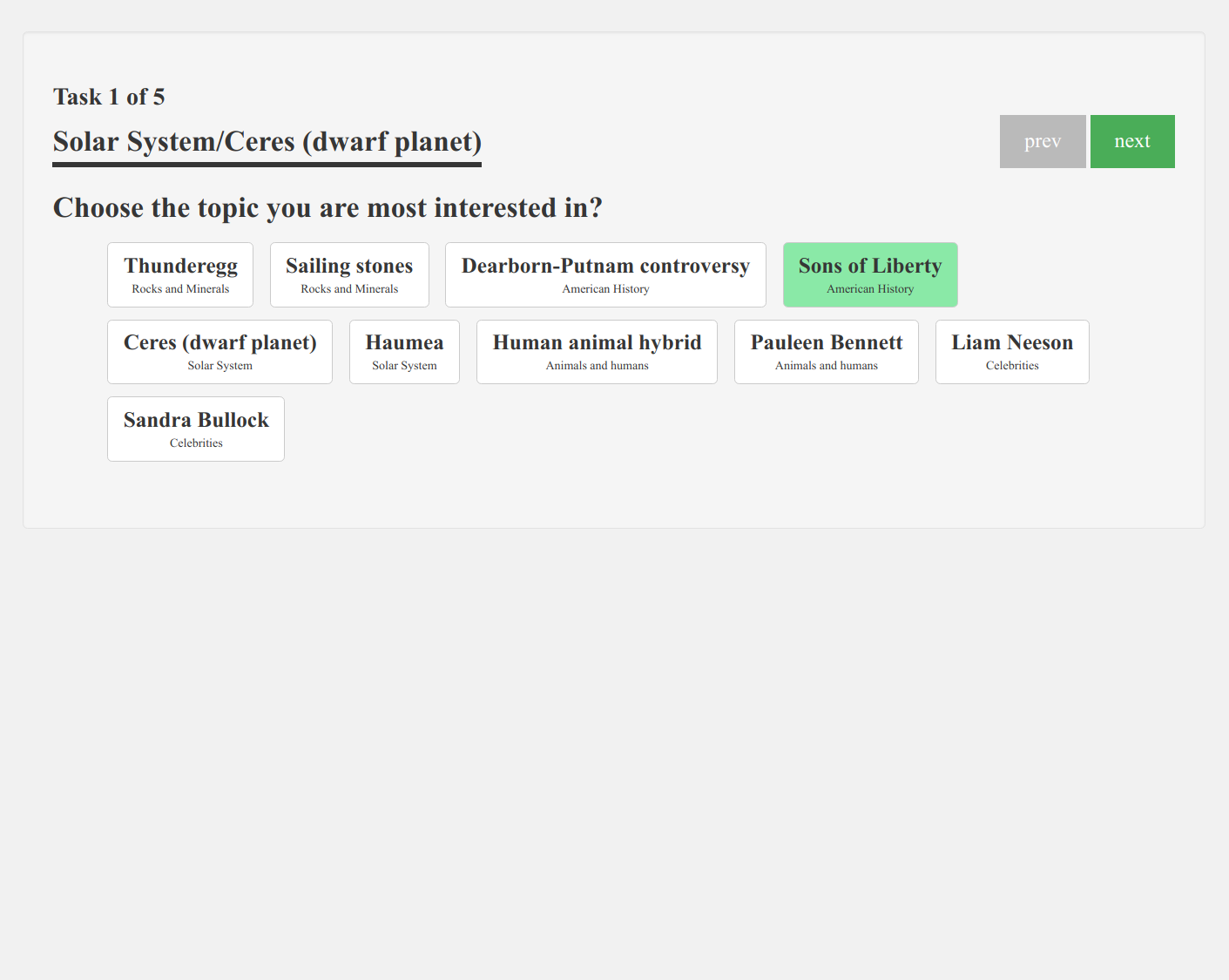}}
\label{fig:screen1}
}\hfill
\subfigure[Step 2: Users enter tags to report their baseline knowledge on the topic.]{
\fbox{\includegraphics[height=4cm]{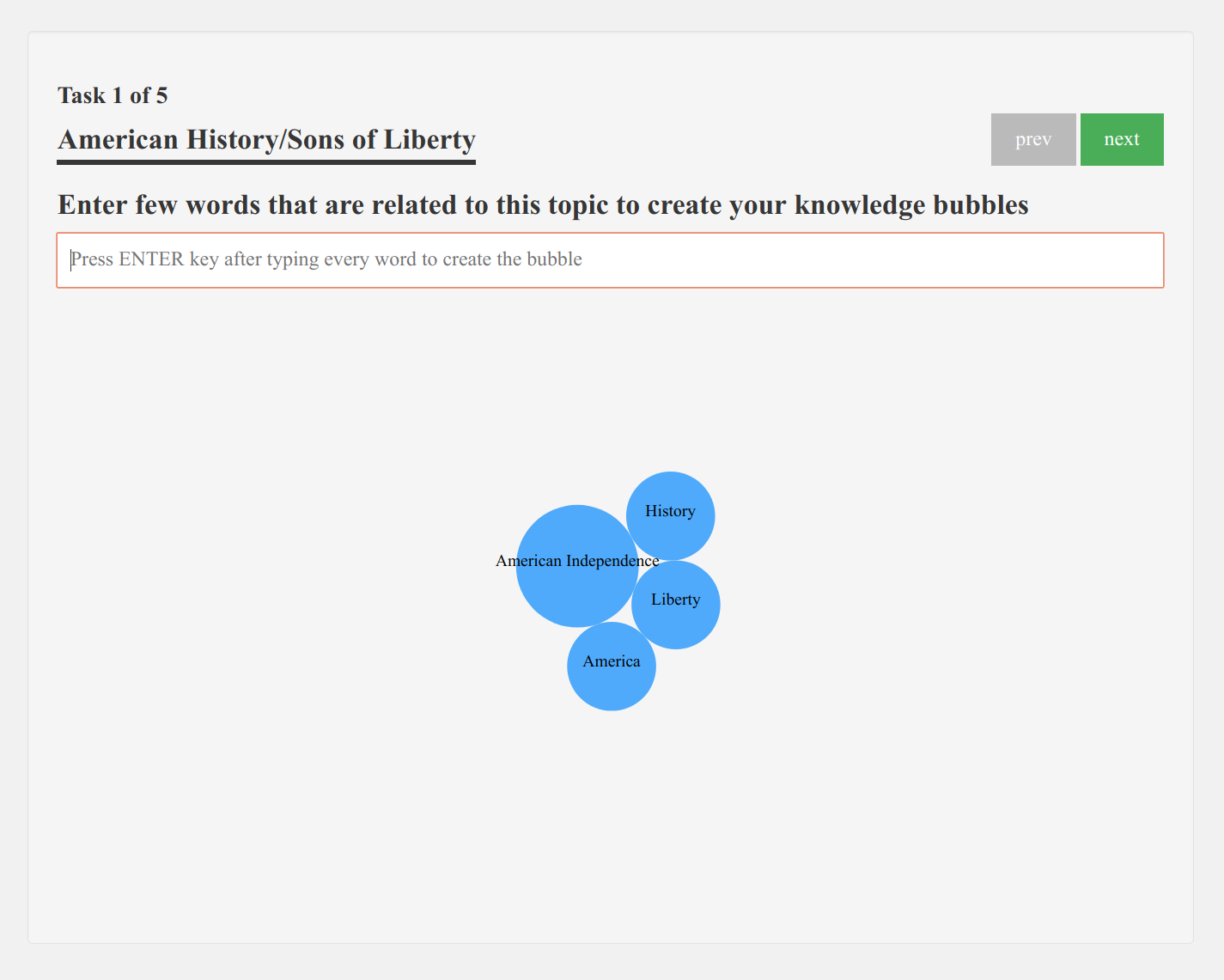}}
\label{fig:pickup}
}\hfill
%\subfigure[Step 3]{
%\fbox{\includegraphics[height=2cm]{figures/steps/screen3}}
%\label{fig:screen2}
%}
\subfigure[Step 3: Users report whether they find a given fact surprising/not-surprising and whether they previously know about that fact.]{
\fbox{\includegraphics[height=4cm]{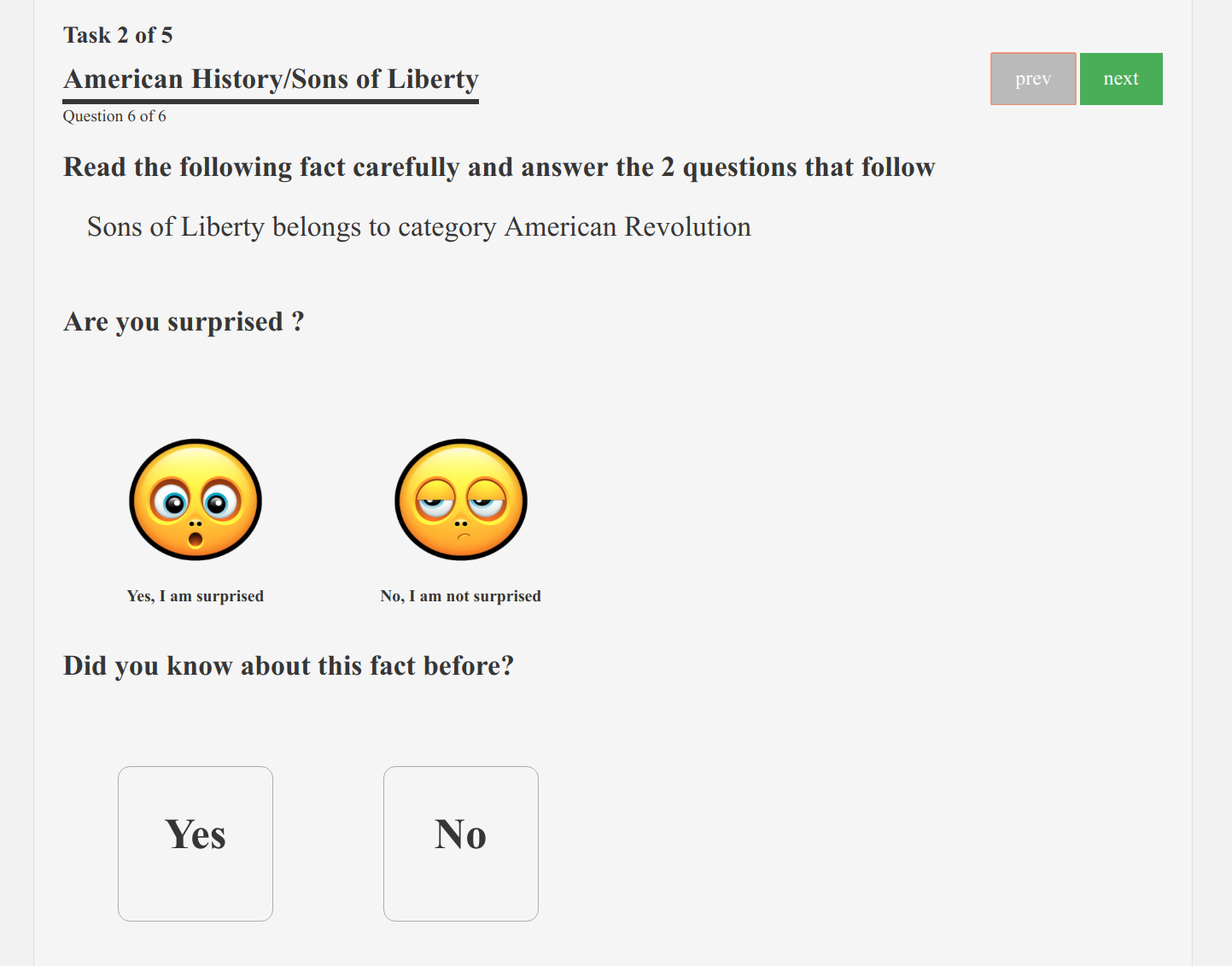}}
\label{fig:decision}
}\hfill
\subfigure[Step 4: Users compare their order of surprisingness with two alternative orders computed by algorithm.]{
\fbox{\includegraphics[height=4cm]{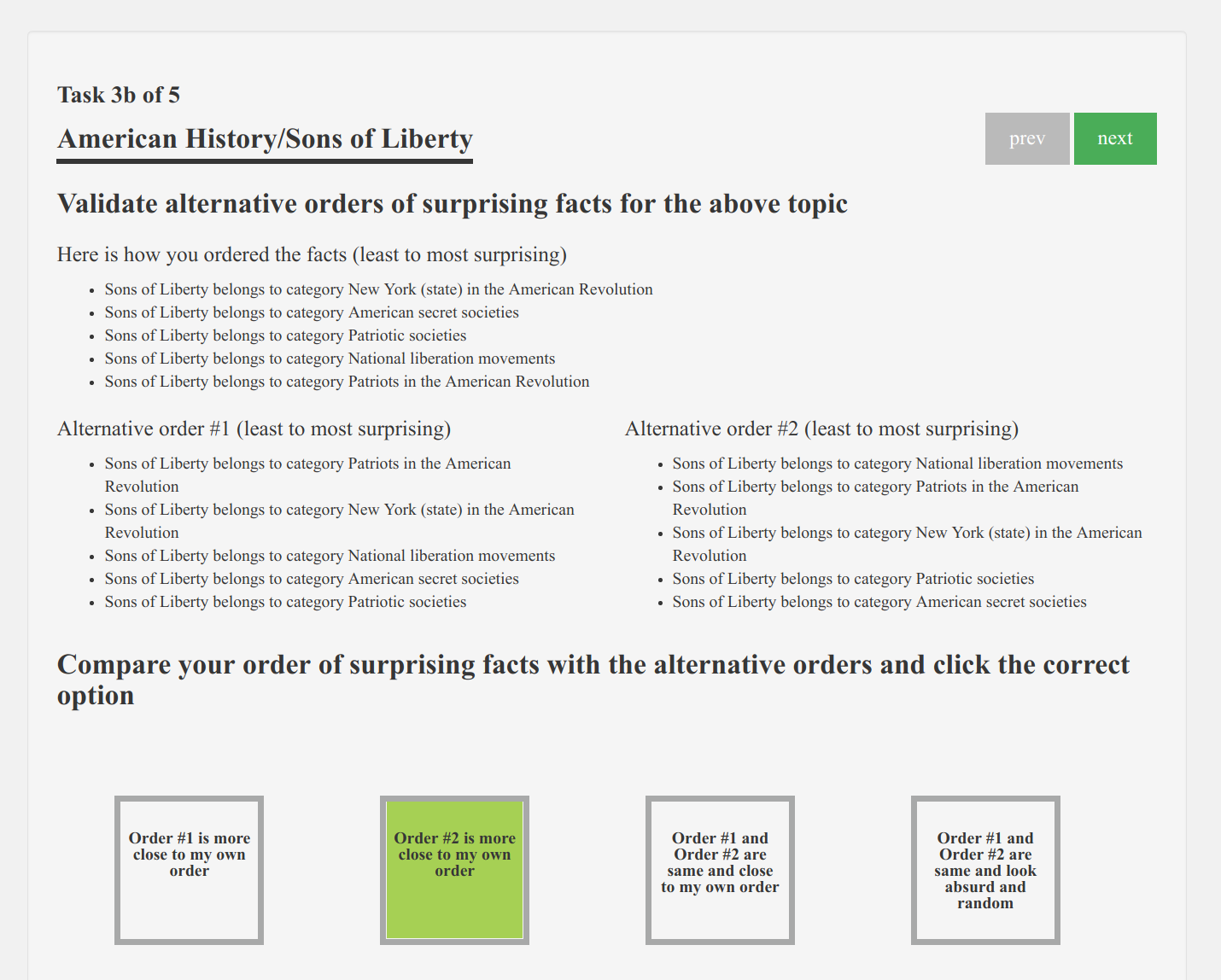}}
\label{fig:compare}
}\hfill
\subfigure[Step 5: Users highlight sentences that they find surprising.]{
\fbox{\includegraphics[height=4cm]{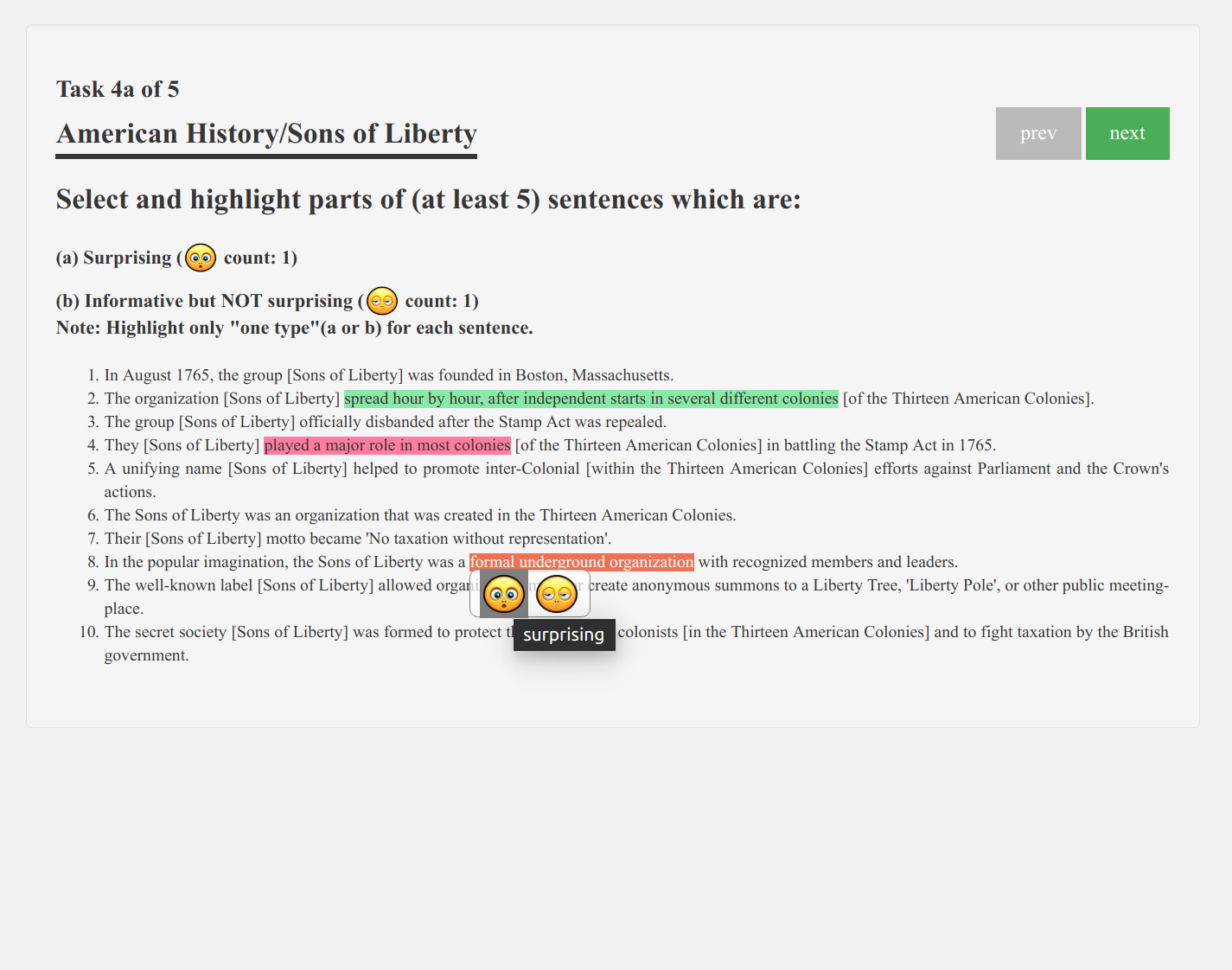}}
\label{fig:select}
}\hfill
\subfigure[Step 6: Users answer questions on how can they change the surprisingness of selected sentences.]{
\fbox{\includegraphics[height=4cm]{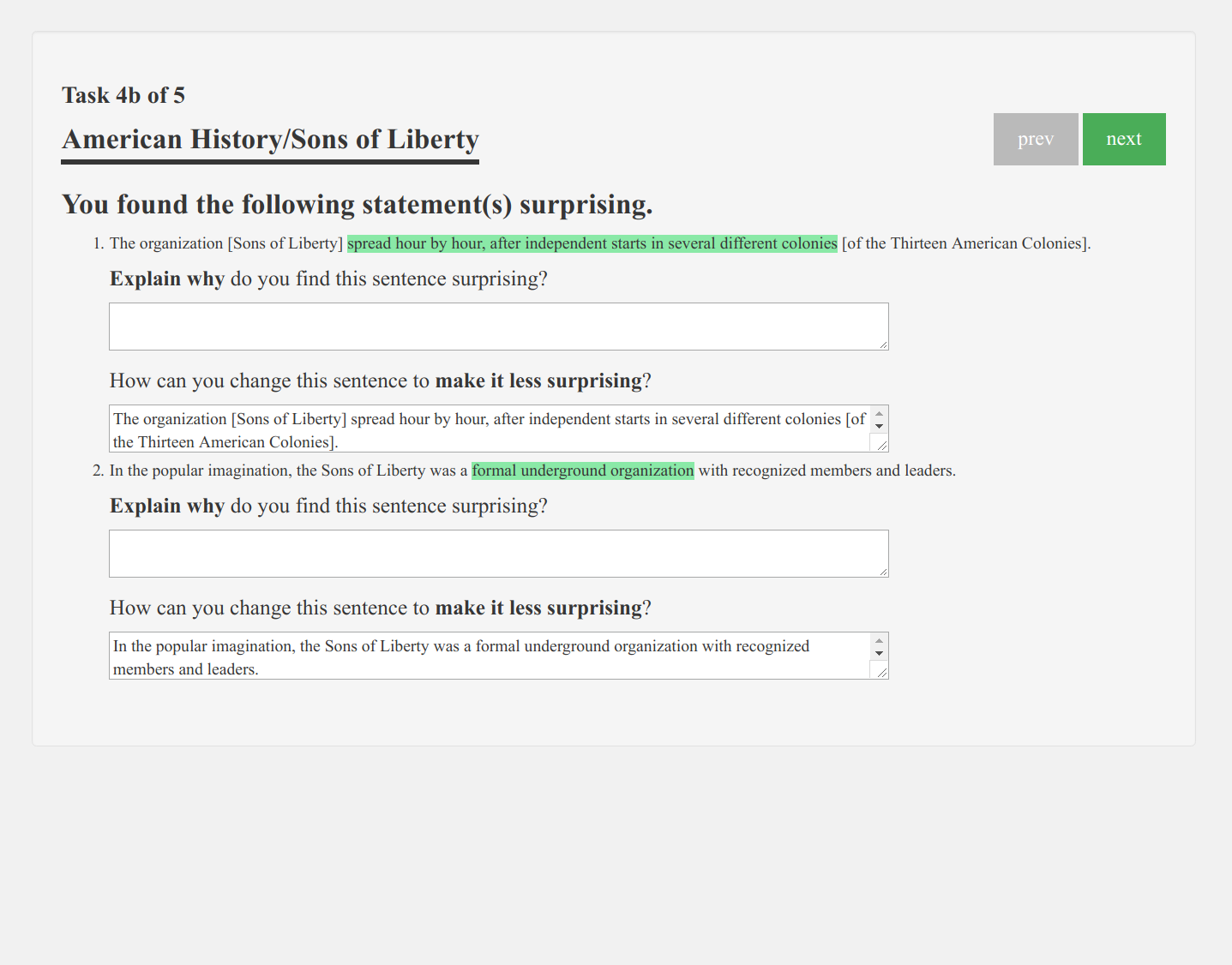}}
\label{fig:explain}
}\hfill
\caption{User Study Task Workflow}
\label{fig:interactions}
%\vspace{-0.5cm}
\end{figure*}

We conducted a user study on Amazon Mechanical Turk with 106 participants, consisting of three parts.  Part 1 of the study asks participants to select an entity (out of $N$ items) that they are most interested in (Figure \ref{fig:screen1}), and to provide tags related to the entity to indicate their level of knowledge about that entity (Figure \ref{fig:pickup}). In Part 2, participants are presented with 15 facts related to that entity of the form ``[entityName] belongs to the category [categoryName]".  For each fact, they are asked whether the fact is surprising or not, and whether they know about the fact previously (Figure \ref{fig:decision}).  Out of those 15 facts, they are given a randomly chosen subset of 5 and asked to rank them based on surprisingness, and to compare their preferred order with two alternative orders produced by our two algorithms (Figure \ref{fig:compare}), without knowing that these are algorithmically generated. In Part 3, participants were given a short paragraph and asked to highlight parts of sentences that they found surprising (Figure \ref{fig:select}), and to describe what changes they would make to the text to increase or decrease the level of surprisingness (Figure \ref{fig:explain}). Each participant received a total of \$3 to complete the task.

\section{Research Questions and Hypotheses}
We explored the following research questions and hypotheses.\\\\
{\bf RQ1: Can computational models of surprise match users' perceptions, when ranking facts by their surprisingness?} \\
\textbf{[H1a]} More users agree with the surprisingness rankings proposed by the computational surprise models.\\
\textbf{[H1b]} Among the participants who thought the rankings suggested by provided are close to their perspective, there is no preference as to which technique is better.\\\\
\noindent {\bf RQ2: How does a participant's knowledge of a fact relate to its surprisingness?} \\
\textbf{[H2]} Participants with knowledge of a certain fact will find the particular fact to be less surprising.

%\begin{table*}[t!]
%\vspace{0.2cm}
 % \centering
%  \caption{Changing Surprising Sentences to be Less Surprising}~\label{tab:more2less}
%  \vspace{-0.3cm}
%  {\scriptsize
%  \begin{tabular}{p{8cm} p{8cm} }
%  \toprule
%   original sentence & modified sentence\\
%   \midrule
%   incorrect-close (IC) & an incorrect but believable response, such as an object that shares at least two attributes in common with what is presented, in terms of function, color, shape, size or material & paper recognized as ``napkin'' \\
%   \midrule
%   incorrect-wayoff (IW) &  a severely incorrect response, e.g., an object that shares no attributes in common with the presented object in terms of function, color, shape, size of material\\
%  \bottomrule
%  \end{tabular}}
%\end{table*}

\section{Results}
There were 106 participants, but not all completed the study. Participants were allowed to answer or skip a question at anytime. Most participants answered the closed ended questions, except for four participants who did not answer any question related to how surprised they were by the facts shown to them. Our statistical models rely on the measurement of surprise and hence, the responses from the aforementioned four participants were not considered for analysis and removed from the database.\\

%The measurements surrounding RQ1 are proportions. The interest is to investigate the proportion of a group of participants is different from the proportion of another group of participants. With the relatively large sample size ($n=102$), the proportion test \cite{agresti2003} was used to answer both \textbf{H1a} and \textbf{H1b}. 

\noindent \textbf{Comparison of Algorithmic and Human Ranking of Surprise}

Among the 85 users who provided feedback on the algorithms, 16 users (19\%) considered both algorithms to be absurd whereas 69 users (81\%) felt that at least one of the algorithms was close to their perspective. Through a one-sided proportion test, we found that there were significantly less users who considered the results from the algorithms to be completely absurd, $z=31.81, p<0.001$. This result shows that at least one of the techniques produces results that are close to the users' perspective. Then our question turns to, which one of these techniques is closer? There were 35 participants (56\%) who thought the ranking from the Technique 1 was closer to their perspective, whereas 28 participants (44\%) thought otherwise. A two-sided proportion test showed that the difference in these two percentages are not statistically significant, $z= 0.57, p = 0.45$, indicating that the rankings produced by both algorithms were equally acceptable to the users.

% is the math for the percentages above correct? Wouldn't 35 participants out of the set of 69, be 51%?

% (omit for space reasons) For the first algorithm, we compared user judgments of <article,category> pair surprise with the algorithm's computed score. We found a strong Spearman's rank correlation ($r=0.27,p=0.015$) between these measures.

%.. original text. reduced to save space.
%our study includes topics of science, history and celebrities whereas they only used celebrities.
\noindent For Technique 1, we found a significant Spearman's rank correlation ($r=0.27,p=0.02$) between averaged user judgments of <article,~category> pair surprise, and the algorithm's computed score. This matches the positive results obtained by Tsurel et al. \cite{tsurel2017fun} on a different evaluation topic (people). Given that our study is based on two additional topics (science, history), our results suggest Technique 1 could generalize well to multiple domains. We also split the <article,~category> pair data by median computed score and found that pairs ranked below the median had nearly half the user-rated surprise ($M=0.27$) compared with those above the median ($M=0.49$). The rated surprise in these two subsets were significantly different by the Kolmogorov-Smirnov test ($p<0.001$). 
We also hypothesize that pairs where the category is less known than the article will be more surprising. We computed \emph{relative familiarity} as the normalized difference between the category term's Google search count and that of the article terms. Adding this to the Technique 1 score yielded an improvement in correlation from ($r=0.27$ to $r=0.33$), suggesting relative familiarity may be another important component in modeling surprise. 
\\

\noindent \textbf{Surprise depends on prior knowledge and assumptions}\\
The participants were asked whether they were surprised each time a fact was presented to them. Their binary answer (Yes/No) is the dependent variable to understand whether knowledge has an impact on surprise. A logistic regression model would have been appropriate \cite{agresti2003} when the dependent variable is binary. However, the answers provided by each participant are correlated within themselves due to unmeasured factors such as education level, occupations and hobbies. To account for the correlation within each participant, the generalized linear mixed effect model is preferred \cite{agresti2003}. The participants are included in the model as random effect and their prior knowledge of each fact is included in the model as an independent variable.

Participants who claimed that they have knowledge of a fact are less likely to claim that they were surprised with the corresponding fact, $\hat{\beta} = -4.09, t(803)=-11.21, p < 0.001$. Participants were not surprised if they already knew the fact or if they had a prior assumption which aligned with the new information. For instance, some participants did not find the fact that \textit{"Sandra Bullock was named 'Most Beautiful Woman' by People magazine in 2015"} surprising. When asked why did they not find this statement surprising, one participant quoted \textit{"I remember seeing this in the magazine"} whereas others said things like \textit{"She just seems to be a beautiful person"} and \textit{"She is obviously beautiful"}. On the other hand, users' were more likely to get surprised with a fact if their prior perception were not aligned with the new information. For instance, one participant found the same fact surprising and said \textit{"[..] did not know this before"}, whereas another participant mentioned \textit{"[..] her age factor"} to be the reason of their surprise. These results implies that prior knowledge and individual assumptions, can play an important role in defining what makes a statement surprising. Thus, algorithmically extracted surprising facts should be used after sensing the baseline knowledge of users.\\ 

\noindent \textbf{Surprise can sometimes be universal}\\
Despite individual differences, we observed from the rankings that some facts were consistently surprising (or unsurprising) for most of the participants. For instance, 10 out of 13 participants (77\%) ranked "Furry fandom" as the most surprising entity belonging to the topic "Human animal hybrid". Likewise, 8 out of 11 participants (73\%) who attempted the topic "Ceres (dwarf planet)", found the fact "Ceres (dwarf planet) belongs to category `named minor planets'" least surprising.\\

\noindent \textbf{Manipulating Surprisingness of Sentences}\\
In part 3, workers were asked to highlight sentences that are surprising, and then make changes to make a surprising sentence less surprising.  Many participants chose to negate the sentence (e.g., add the word ``not"). Other participants took the approach of adding content to make the sentence more contradictory, or removing the surprising content.    For instance, when asked to make the sentence \textit{"Empire magazine ranked Liam Neeson among both the `100 Sexiest Stars in Film History' and `The Top 100 Movie Stars of All Time'"} less surprising, participants reduced the sentence to "Empire magazine ranked Liam Neeson in two of its polls", making it more general. Similarly, when asked to increase the surprisingness of the fact "[Sandra Bullock] was named `Most Beautiful Woman' by People magazine in 2015.", participants added new information and changed the sentence to "[..] and beat Jennifer Lopez."  This part of the user study failed to produce any meaningful results---in general, the changes to the sentences were minimal (e.g., 1-2 words) or semantically irrelevant (i.e., did not change the meaning of the original sentence at all). In future work, we will develop a better design for this {\it generative surprise} task to probe participants' mental models of surprise.   One participant mentioned that it ``seems difficult to make [the sentence] more surprising without more information."  A possible direction is to create a mixed-initiative surprise generation system that recommends information to add/remove that can make a sentence more or less surprising.

\section{Conclusions}
In this paper, we report findings from a user study where participants are asked to rate and rank the surprisingness of facts, and compare their ranking of surprise against those generated by algorithms.   Results show that both algorithms we investigated perform well in terms of finding facts that are also perceived to be surprising by users.  Qualitative analysis reveal that surprise strongly depends on individual differences in prior knowledge and assumptions; at the same time, some facts can also be generally surprising or unsurprising to the crowd.